\documentclass[twocolumn,preprintnumbers,showpacs,showkeys,amsmath,amssymb,floatfix]{revtex4}
\usepackage{graphicx}
\usepackage{dcolumn}
\usepackage{bm}
\begin{document}
\title{Repelling Random Walkers in a Diffusion-Coalescence System}
\author{F. H. Jafarpour $^{1}$}
\email{farhad@ipm.ir}
\author{S. R. Masharian$^2$}
\affiliation{$^1$ Physics Department, Bu-Ali Sina University, Hamadan, Iran}
\affiliation{$^2$ Institute for Advanced Studies in Basic Sciences, Zanjan, Iran}
\date{\today}
\begin{abstract}
We have shown that the steady state probability distribution
function of a diffusion-coalescence system on a one-dimensional
lattice of length $L$ with reflecting boundaries can be written in
terms of a superposition of double shock structures which perform
biased random walks on the lattice while repelling each other. The
shocks can enter into the system and leave it from the boundaries.
Depending on the microscopic reaction rates, the system is known to
have two different phases. We have found that the mean distance
between the shock positions is of order $L$ in one phase while it is
of order $1$ in the other phase.
\end{abstract}
\pacs{02.50.Ey, 05.20.-y, 05.70.Fh, 05.70.Ln}
\keywords{Reaction-Diffusion Systems, Random Walk, Shock}
\maketitle
\begin{figure}
\includegraphics[width=3in]{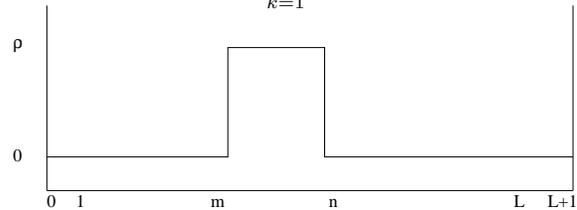}
\caption{\label{fig} Sketch of a double shock structure. The shock
positions are defined at the sites $m$ and $n$..}
\end{figure}

Recently the investigation of the microscopic structure and dynamics
of shocks defined as discontinuities in the space dependence of the
densities of particles in one-dimensional driven diffusive systems,
has drawn much attention \cite{FER}-\cite{AA2}. It has been shown
that the steady states of some of these systems can be explained in
terms of collective excitations with one or more conservation laws. \\
In \cite{KJS} three families of single-species driven-diffusive
systems are studied in which a traveling shock with a step-like
density profile exists and behaves like a one-particle excitation in
the system provided that the microscopic hopping rates are fine
tuned. This has also been observed in the systems with more than one
species of particles \cite{JM1}-\cite{TS2}. On the other hand, the
steady states of these systems can be written in terms of a
superposition of such product shock measures. In \cite{JM2} and
\cite{J2} the authors have shown that such steady states are
associated with the existence of two-dimensional representations of
the quadratic algebras of these systems when they are studied using
the Matrix Product Formalism (MPF) (for a recent review see
\cite{BE}). According to this formalism the steady state of some of
one-dimensional driven-diffusive systems can be written in terms of
products of noncommuting operators which satisfy a quadratic algebra.\\
However, little is known about the microscopic dynamics of multiple
shocks in these systems. The only example is given in \cite{BS}
where multiple shocks are studied for the partially asymmetric
simple exclusion process with open boundaries. In this paper we
investigate the dynamics of a double shock structure in a branching
coalescing system with nonconserving dynamics and reflecting
boundaries. The steady state properties of this system has already
been studied in \cite{HKP} and \cite{HSP}. It turns out that
depending on the microscopic reaction rates of the system it can be
in two different phases: a high-density and a low-density phase.
Since the dynamics of the system is non-conserving, the mean density
of the particles in the system in high-density phase is greater than
that in the low-density phase. However, it has been shown that if
one considers a canonical ensemble in which the total number of
particles is conserved then the system has two phases: a
high-density and a shock phase. In this case the shock does not have
any dynamics \cite{JM3}. In \cite{KJS} the authors have shown that a
single shock with biased random walk dynamics can evolve in this
system provided that the boundaries are open so that the particles
can enter and leave the system from there. Later in \cite{AA1} and
\cite{AA2} it was shown that in an infinite system double shock
structures with random walk dynamics can also evolve in the system.
However, nothing is known about the dynamics of these double shock
structures in a system with boundaries. Our main attempt in this
paper is to study the microscopic dynamics of such structures on
a lattice of finite length and reflecting boundaries. \\
In what follows we first define the model and then using the
Hamiltonian formalism show how a double shock product measure
evolves in time under the Hamiltonian of the system. From there we
construct the steady state probability distribution function of the
system as a linear combination of such double product shock
measures. The mean distance between the shock positions is also
calculated in the thermodynamic limit. \\
The system in question consists of identical classical particles on
a one-dimensional lattice of length $L$. There is no injection or
extraction of particles at the boundaries. The reaction rules
between two consecutive sites $k$ and $k+1$ on the lattice are as
follows:
\begin{equation}
\label{Rules}
\begin{array}{ll}
\emptyset+A \rightarrow A+\emptyset & \mbox{with rate} \; \; q \\
A+\emptyset \rightarrow \emptyset+A & \mbox{with rate} \; \; q^{-1} \\
A+A \rightarrow A+\emptyset & \mbox{with rate} \; \; q \\
A+A \rightarrow \emptyset+A & \mbox{with rate} \; \; q^{-1} \\
\emptyset+A \rightarrow A+A & \mbox{with rate} \; \; \Delta q \\
A+\emptyset \rightarrow A+A & \mbox{with rate} \; \; \Delta q^{-1}\\
\end{array}
\end{equation}
in which $A$ and $\emptyset$ stand for the presence of a particle
and a hole respectively. As can be seen, the parameter $q$
determines the asymmetry of the system. For $q>1$ ($q<1$) the particles
have a tendency to move in the leftward (rightward) direction. For any $q$
the model is also invariant under the following transformations:
\begin{equation}
\label{SY} q\longrightarrow q^{-1} \;,\; k\longrightarrow L-k+1
\end{equation}
in which $k$ is a given site on the lattice. Throughout this paper
we will only consider the case $q > 1$. The results for the case $q
< 1$ can easily be obtained using (\ref{SY}). By formulating the
stochastic Hamiltonian of the system as a quantum spin chain, it has
been shown that it is completely integrable \cite{HKP,HSP}. As we
mentioned the system has two different phases depending on the
values of $q$ and $\Delta$. For $q > 1$ it has a high-density phase
for $q^2 < 1+\Delta$ and a low-density phase for $q^2 > 1+\Delta$.
In the high-density phase the density profile of particles has its
maximum value near the left boundary while it is a constant
$\rho=\frac{\Delta}{1+\Delta}$ in the bulk of the lattice. It also
drops exponentially to zero near the right boundary. The particle
correlations exist at both boundaries. In the low-density phase the
density profile of particle has again its maximum value near the
left boundary but it quickly drops exponentially to zero in the bulk
and remains zero throughout the lattice. In this phase the particle
correlations only exist near the left boundary. The mean density of
particles is of order $\frac{1}{L}$ in this phase. On the transition
line $q^2=1+\Delta$ the density profile of particles drops
exponentially near the left boundary while changes linearly in the
bulk of the system. The mean density of particles in the bulk of
the lattice is equal to $\frac{\Delta}{2(1+\Delta)}$ in the thermodynamic
limit.\\
Recently, it has been shown that the steady state probability
distribution function of some of one-dimensional driven-diffusive
systems can be written in terms of interactions of single shock
structures \cite{JM2}. In the following we will show that the steady
state of our coalescence system defined by (\ref{Rules}) can also be
written in terms of superposition of double shock structures. These shocks
repel each other while perform biased random walk on the lattice.\\
Any state of the system is defined through a probability measure
$P_\eta$ on the set of all configurations $\eta=(\eta_1,\eta_2,
\dots,\eta_L)$, $\eta_k\in\{\emptyset, A \}$. For our purposes it is
convenient to use the Hamiltonian formalism where one assigns a
basis vector $\vert \eta\rangle$ of the vector space
$(\mathbb{C}^2)^{\otimes L}$ to each configuration and the
probability vector is defined by $\vert P\rangle=\sum_\eta P_\eta
\vert \eta\rangle$ which is normalized such that $\langle s \vert
P\rangle=1$ where $\langle s \vert =\sum_\eta \langle\eta \vert $.
The time evolution is now described by the master equation:
\begin{equation}
 \frac{d}{dt} \vert P(t) \rangle = H \vert P(t) \rangle
\end{equation}
in which $H$ is called the Hamiltonian and its matrix elements are
the hopping rates between any two configurations. For a system
defined on a lattice of length $L$ with reflecting boundaries the
Hamiltonian can be written as:
\begin{equation}
\label{H}
H=\sum_{k=1}^{L-1} h_{k,k+1},
\end{equation}
where $h_{k,k+1}$ acts nontrivially only on sites $k$ and $k+1$. In
a basis defined as:
\begin{equation}
  \vert \emptyset\rangle =
  \left(\begin{array}{c}
    1 \\ 0
  \end{array}\right), \quad
 \vert A\rangle =
  \left(\begin{array}{c}
    0 \\ 1
  \end{array}\right)
\end{equation}
the local Hamiltonian of our system in (\ref{H}) has the following form:
\begin{equation}
h_{k,k+1}=\left(
     \begin{array}{cccc}
        0& 0& 0& 0\\
        0& -q(1+\Delta)& q^{-1}& q^{-1}\\
        0& q& -q^{-1}(1+\Delta)& q\\
        0& q\Delta& q^{-1}\Delta& -(q+q^{-1})\\
        \end{array}
  \right).
\end{equation}
We define a double Bernoulli shock measure which is a product
measure with two jumps in the local particle density associated with
two random walkers (the shock fronts) at sites $m$ and $n$ as:
\begin{equation}
  \label{P_{m,n}}
  \vert P_{m,n} \rangle =
  \left(\begin{array}{c}
    1 \\ 0
  \end{array}\right)^{\otimes m} \otimes
  \left(\begin{array}{c}
    1-\rho \\ \rho
  \end{array}\right)^{\otimes n-m-1}
\otimes
  \left(\begin{array}{c}
    1 \\ 0
  \end{array}\right)^{\otimes L-n+1}
\end{equation}
in which $0 \leq m \leq n-1$ and $1 \leq n \leq L+1$. Here we have
introduced two auxiliary sites $0$ and $L+1$. A simple sketch of
such shock measure is given in FIG \ref{fig}.
It is easy to verify that this family of shock measures generate a
subspace of the vector-space of states which is invariant under the
time evolution generated by $H$ and thus the many-particle problem
is reduced to a two-particle one. As we mentioned earlier, the time
evolution of such product shock measure has already been studied for
an infinite system with no boundaries \cite{AA1,AA2}; nevertheless,
in this paper we aim to study a finite system with reflecting
boundaries. The time evolution equations for $\vert P_{m,n}\rangle$
are given by:
\begin{widetext}
\begin{equation}
\label{Equs}
\begin{array}{l}
H \vert P_{m,n} \rangle = q^{-1} \vert P_{m+1,n} \rangle + q(1+\Delta) \vert P_{m-1,n} \rangle
+q^{-1}(1+\Delta)\vert P_{m,n+1} \rangle+q \vert P_{m,n-1}
\rangle -(q+q^{-1})(2+\Delta)\vert P_{m,n} \rangle \quad \mbox{for}\\ \\
m=1,\cdots,L-2 \quad \mbox{and} \quad n=m+2,\cdots,L\\ \\
H \vert P_{0,n} \rangle=(q^{-1}-q) \vert P_{1,n} \rangle+q^{-1}(1+\Delta)
\vert P_{0,n+1} \rangle+q  \vert P_{0,n-1} \rangle
-q^{-1}(2+\Delta)\vert P_{0,n} \rangle \quad \mbox{for} \quad n=2,\cdots,L\\ \\
H \vert P_{m,L+1} \rangle=q^{-1}\vert P_{m+1,L+1} \rangle+q(1+\Delta)\vert
P_{m-1,L+1} \rangle+(q-q^{-1})\vert P_{m,L} \rangle-
q(2+\Delta)\vert P_{m,L+1} \rangle \quad \mbox{for} \quad m=1,\cdots,L-1\\ \\
H\vert P_{0,L+1} \rangle=(q^{-1}-q)\vert P_{1,L+1} \rangle+(q-q^{-1})\vert P_{0,L} \rangle\\ \\
H\vert P_{m,m+1} \rangle=0 \quad \mbox{for} \quad m=0,\cdots,L. \\
\end{array}
\end{equation}
\end{widetext}
As can be seen for $q > 1$ the left random walker performs a biased random
walk and preferentially hops to the left regardless of the values of $q$ and
$\Delta$. In contrast the right random walker preferentially hops to the left
for $q^2 > 1+\Delta$  and to the right for $q^2 < 1+\Delta$. On the coexistence
line where $q^2=1+\Delta$ the right random walker performs an unbiased random walk.
The left (right) random walker can also leave the lattice only from the left (right)
boundary. The diffusion coefficients and also the velocities of the random walkers
can now be easily calculated from (\ref{Equs}). \\
Let us now explain why the random walkers repel each other. It can
easily be seen from (\ref{Equs}) that as long as the shock positions
are more than a single site apart, they never meet each other during
the time evolution. However, it seems from there that the random
walkers can meet each other when they are a single site apart. In
what follows we show that this is not the case. For instance we
consider the first equation in (\ref{Equs}) for $n=m+2$ where the
shock positions are a single site apart. Rewriting this equation in
terms of a new definition for the shock measure as:
\begin{equation}
  \vert {\tilde P}_{m,n} \rangle =
  \left(\begin{array}{c}
    1 \\ 0
  \end{array}\right)^{\otimes m} \otimes
  \left(\begin{array}{c}
    0 \\ 1
  \end{array}\right)^{\otimes n-m-1}
\otimes
  \left(\begin{array}{c}
    1 \\ 0
  \end{array}\right)^{\otimes L-n+1}
\end{equation}
one finds:
\begin{widetext}
$$
H \vert {\tilde P}_{m,m+2} \rangle = q\Delta \vert {\tilde
P}_{m-1,m+2} \rangle + q \vert {\tilde P}_{m-1,m+1} \rangle
+q^{-1}\vert {\tilde P}_{m+1,m+3} \rangle+q^{-1} \Delta \vert
{\tilde P}_{m,m+3} \rangle -(q+q^{-1})(1+\Delta)\vert {\tilde
P}_{m,m+2} \rangle.
$$
\end{widetext}
As can be seen the shock positions never get closer that a singe
site. In fact the dynamical rules (\ref{Rules}) do not allow the
shock fronts to get closer than a single site since it results in an
empty lattice. This is why we say that the random walkers repel each
other. One can easily check this for other equations in (\ref{Equs})
in which the shock positions are a single site apart to see that in
terms of the $\vert {\tilde P}_{m,n} \rangle$ the shock positions
never meet and the minimum distance between them is at least a
single site. \\
In this paper we are specially interested in the steady state of the
system. One should note that an empty lattice is a trivial steady
state for the system. It can be seen from (\ref{Rules}) that an
empty lattice never evolves in time. There is actually a nontrivial
steady state for the system in which the lattice contains some
particles. The nontrivial steady state of the system can now be
constructed as a superposition of double shock measures as follows:
\begin{equation}
\label{SS}
\vert P^{*} \rangle=\frac{1}{Z_L}\sum_{m=0}^{L}\sum_{n=m+1}^{L+1} \psi_{m,n}\vert P_{m,n} \rangle
\end{equation}
provided that we exclude the empty lattice from $\vert P^{*} \rangle$ by requiring:
\begin{equation}
\label{Cond}
 \langle 0 \vert P^{*} \rangle=0
\end{equation}
in which:
\begin{equation}
 \vert 0 \rangle=\vert \emptyset \rangle ^{\otimes L}=\left(
     \begin{array}{c}
        1\\
        0\\
        \vdots\\
        0\\
        \end{array}
  \right)
\end{equation}
is associated with a configuration with no particles in the system.
The normalization factor $Z_L$ in (\ref{SS}) can easily be obtained
from $Z_L=\sum_{m=0}^{L}\sum_{n=m+1}^{L+1}\psi_{m,n}$. By requiring
$H \vert P^{*} \rangle =0$ we find a system of equations for
$\psi_{m,n}$'s. It turns out that this system of equations has the
following solution:
\begin{widetext}
\begin{equation}
\label{Solu}
\psi_{m,n} =\frac{(q^2-1)}{(1-q^2)^{\delta_{m,0}}(\frac{q^2-1}{q^2})^{\delta_{n,L+1}}}
q^{-2(m+n)}(1-(1+\Delta)^{n-m-1}) \quad
\mbox{for} \; 0 \leq m \leq L-1 \; \mbox{and} \; m+2 \leq n \leq L+1.
\end{equation}
\end{widetext}
Note that $\psi_{m,n}$'s in (\ref{Solu}) are also valid for $q^2 =
1+\Delta$ and at the coexistence line one should only replace
$\Delta$ with $q^2-1$ in (\ref{Solu}). One can see from
(\ref{P_{m,n}}) that there are $L+1$ states in which the shock
positions are at two consecutive sites. The states $\vert
P_{m,m+1}\rangle$'s point to an empty lattice.
Since the empty lattice is a trivial steady state of the system;
therefore, the coefficient of these states in (\ref{SS}) i.e.
$\psi_{m,m+1}$'s, are taken to be equal to $\psi'$. The condition
(\ref{Cond}) for $q^2\neq 1+\Delta$ can now be calculated and it is
equal to:
\begin{equation}
\label{psi1}
\begin{array}{ll}
\psi' & =\frac{1}{L+1}[\frac{q^2\Delta}{(q^4-1)(1-q^2(1+\Delta))}+\\ \\
& \frac{q^2\Delta^2}{(q^2-1)(q^2-1-\Delta)(1-q^2(1+\Delta))}(\frac{1}{q^2(1+\Delta)})^L+\\ \\
& \frac{q^2\Delta}{(q^4-1)(q^2-1-\Delta)}q^{-4L}]. \\ \\ \\ \\
\end{array}
\end{equation}
It turns out that on the transition line $q^2=1+\Delta$, $\psi'$ becomes:
\begin{equation}
\label{psi2}
\psi'=-\frac{((q^4-1)L-q^2)q^{-4L}+q^{2}}{(L+1)(q^4-1)(q^2+1)}.
\end{equation}
The normalization factor $Z_L$ which is called the grand-canonical
partition function of the system can now be calculated and after
substituting $\psi'$ from (\ref{psi1}) and (\ref{psi2}) one finds:
\begin{widetext}
\begin{equation}
\label{PF}
Z_L=\left\{
\begin{array}{ll}
\frac{q^2\Delta^2(q^2-1)^{-1}}{(1-q^2(1+\Delta))(1+\Delta-q^2)}[1-(\frac{1+\Delta}{q^2})^L-
(\frac{1}{q^2(1+\Delta)})^L+q^{-4L}]&\mbox{for $q^2 \neq 1+\Delta$}\\ \\
\frac{1-q^{-4L}}{1+q^2}L&\mbox{for $q^2 = 1+\Delta$}.
\end{array}
\right.
\end{equation}
\end{widetext}
As one can see our results obtained here are exactly those obtained
in \cite{HKP} and \cite{HSP} using different approaches. Using the
steady state probability distribution function (\ref{SS}) one can
easily calculate the density profile of the particles and also any
correlations in the steady state. However, since the results are
exactly those obtained in the above mentioned papers,
the results are not given here.\\
Having the probability of finding the random walkers at sites $m$
and $n$ in the steady state, one can calculate the mean distance of
the shock fronts in the steady state defined as:
\begin{equation}
 \langle d \rangle = \frac{1}{Z_L}\sum_{m=0}^{L}\sum_{n=m+1}^{L+1}(n-m-1)\psi_{m,n}.
\end{equation}
It turns out that in the thermodynamic limit $L \rightarrow \infty$ it is given by:
\begin{equation}
\langle d \rangle \sim \left\{
\begin{array}{ll}
L \quad \mbox{for} \quad q^2 < 1+\Delta\\ \\
\frac{1}{2}L \quad \mbox{for} \quad q^2 = 1+\Delta \\ \\
{\mathcal O} (1) \quad \mbox{for} \quad q^2 > 1+\Delta.
\end{array}
\right.
\end{equation}
In the high-density phase $q^2 < 1+\Delta$ the shock fronts have
their maximum distance while in the low-density phase they have the
minimum distance which is of order of a single site. One should note
that the mean distance of the two shock fronts changes abruptly from
one phase to the other phase which can be a sign for the phase
transition in the system. \\
It is also interesting to study the probability of finding each
shock front at a given site in the steady state. The probability of
finding the left shock front at the site $m$ is defined as:
\begin{equation}
 P_{m}=\sum_{n=m+1}^{L+1}\psi_{m,n} \quad \mbox{for} \quad 0 \leq m \leq L.
\end{equation}
In the thermodynamic limit $L\rightarrow \infty$ and in the
high-density phase $P_{m}$ is an exponential function with the
inverse length scale $\ln (q^4)$ while in the low-density phase it
is an exponential function with the inverse length scale equal to
$\ln (q^2(1+\Delta))$. On the other hand the probability of finding
the right shock front at site $n$ is given by:
\begin{equation}
 P_{n}=\sum_{m=0}^{n-1}\psi_{m,n} \quad \mbox{for} \quad 1 \leq n \leq L+1.
\end{equation}
In the thermodynamic limit $L\rightarrow \infty$ and in both the
high-density and the low-density phase this probability distribution
function has an exponential behavior with the inverse length scale
$\ln (q^{2}(1+\Delta)^{-1})$. This explains
why the system has three diffrent length scales.\\
In this paper we have studied a coalescence system with reflecting
boundaries and showed that its steady state can be explained in
terms of superposition of probability distribution of two
interacting random walkers which perform biased random walks while
repelling each other. The random walkers can also leave or enter
from the boundaries. One should note that the random walk picture
actually fails at the left boundary for $q > 1$ (and at the right
boundary for $q < 1$). In fact, as can be seen from (\ref{Equs}),
the left random walker should enter the system with a negative rate.
This has already been observed in the branching-coalescing model
with open boundaries studied in \cite{KJS}. Apart from this, we have
found that the steady state probability distribution function of the
system is exactly the one obtained in \cite{HKP,HSP} which obviously
generates the same density profile of particles in the system in
each phase as it was calculated by the same authors. It is
interesting to consider a more general reaction rates in
(\ref{Rules}) and see under what constraints the random walk picture
in a system with refelecting boundaries exists. This is under our
investigations and will be published elsewhere.

\end{document}